\documentclass[
journal=ancac3, % for ACS Nano
manuscript=article]{achemso}

\usepackage{siunitx}
\usepackage[english=british]{csquotes}
\usepackage{xcolor}

\usepackage{xr}

\makeatletter
\newcommand*{\addFileDependency}[1]{% argument=file name and extension
  \typeout{(#1)}
  \@addtofilelist{#1}
  \IfFileExists{#1}{}{\typeout{No file #1.}}
}
\makeatother

\newcommand*{\myexternaldocument}[1]{%
    \externaldocument{#1}%
    \addFileDependency{#1.tex}%
    \addFileDependency{#1.aux}%
}

\myexternaldocument{zzSI_arxiv}

\usepackage{afterpage}

\providecommand{\Revised}[1]{#1}

\DeclareSIUnit\Molar{\textsc{m}}

\author{Debora Monego\footnote{These authors contributed equally.}}
\affiliation[University of Sydney]
{ARC Centre of Excellence in Exciton Science, School of Chemistry and The University of Sydney Nano Institute, University of Sydney, Sydney, New South Wales 2006, Australia}

\author{Thomas Kister\textsuperscript{a}}
\affiliation[INM --- Leibniz Institute for New Materials]
{INM --- Leibniz Institute for New Materials, Campus D2 2, 66123 Saarbr\"ucken, Germany}

\author{Nicholas Kirkwood}
\affiliation[University of Melbourne]
{ARC Centre of Excellence in Exciton Science, School of Chemistry, University of Melbourne, Parkville, Victoria 3010, Australia}

\author{Paul Mulvaney}
\affiliation[University of Melbourne]
{ARC Centre of Excellence in Exciton Science, School of Chemistry, University of Melbourne, Parkville, Victoria 3010, Australia}

\author{Asaph Widmer-Cooper}
\affiliation[University of Sydney]
{ARC Centre of Excellence in Exciton Science, School of Chemistry and The University of Sydney Nano Institute, University of Sydney, Sydney, New South Wales 2006, Australia}

\author{Tobias Kraus}
\email{tobias.kraus@leibniz-inm.de}
\affiliation[INM --- Leibniz Institute for New Materials]
{INM --- Leibniz Institute for New Materials, Campus D2 2, 66123 Saarbr\"ucken, Germany\\
Colloid and interface chemistry, Saarland University, Campus D2 2, 66123 Saarbr\"ucken, Germany}

\title[]
{On the colloidal stability of apolar nanoparticles:\\ The role of ligand length} 

\begin{document}

\newpage

\begin{abstract}
Inorganic nanoparticle cores are often coated with organic ligands to render them dispersible in apolar solvents. However, the effect of the ligand shell on the \Revised{colloidal} stability of the overall hybrid particle is not fully understood. In particular, it is not known how the length of an apolar alkyl ligand chain affects the stability of a nanoparticle dispersion against agglomeration.Here, Small-Angle X-ray Scattering and molecular dynamics simulations have been used to study the interactions between gold nanoparticles and between cadmium selenide nanoparticles passivated by alkanethiol ligands with 12 to 18 carbons in the solvent decane. We find that increasing the ligand length increases \Revised{colloidal} stability in the core-dominated regime but decreases it in the ligand-dominated regime. This unexpected inversion is connected to the transition from ligand- to core-dominated agglomeration when the core diameter increases at constant ligand length. Our results provide a microscopic picture of the forces that determine the colloidal stability of apolar nanoparticles and explain why classical colloid theory fails.
\end{abstract}

\section{Introduction}

The most common way to stabilize inorganic nanoparticles in apolar solvents is to coat them with sufficiently dense layers of molecules with apolar chains\cite{batista2015nonadditivity,Boles2016,Si2018}. Nanoparticles made from noble metals (for instance Au, Ag, or Pt) \cite{zheng2006one}, semiconductors (for instance CdSe, CdTe, or PbSe) \cite{murray1993synthesis}, metal oxides (for instance Fe3O4, TiO2, or Al2O3) \cite{park2004ultra}, and alloys (AuAg, AuCu, or FePt) \cite{he2014facile, sun2000monodisperse}, have thus been coated with organic compounds\cite{brust1994synthesis} that lower their interfacial energy and add steric stabilization\cite{napper1977steric}.  Suitable organic compounds include alkanethiols \cite{zheng2006one}, fatty acids \cite{park2004ultra}, other surfactants \cite{kvitek2008effect}, and polymers \cite{corbierre2001polymer}. They require binding groups with sufficient affinity for the nanoparticle surface \cite{grubbs2007roles} that usually contain nitrogen\cite{sanchez2010binary}, oxygen\cite{de2015entropy}, phosphorus\cite{murray1993synthesis}, or sulfur (as in alkanethiols) \cite{zheng2006one}. Coinage metal nanocrystals are often stabilized with alkanethiols or alkylamines \cite{zheng2006one, mourdikoudis2013oleylamine}, less frequently with carboxylic acids, phosphines, or phosphonates. \cite{le2010synthesis, weare2000improved} Conversely, metal oxide nanoparticles are often coated with alkylamines or carboxylic acids, \cite{mourdikoudis2013oleylamine, park2004ultra} while all of the above classes of surfactant have been used to passivate semiconductor nanocrystals.

The resulting \enquote{ligand shells} have similarities to self-assembled monolayers, but the curvature and the typical facets of inorganic nanoparticles complicate the structure and dynamics of the ligand shell. Ligand molecules may bind to different nanocrystal facets at different densities, an effect that is important for the growth of anisotropic nanostructures in solution \cite{Lee2002, Lee2003}, while the high curvature of small particles leads to the \enquote{hairy ball effect}, where the tails of the ligands have access to considerably more free volume than the head groups. Ligand shells are also more dynamic than sometimes envisioned; even simple alkanethiol coatings exhibit phase transitions where the shell changes from a more mobile disordered state to a less mobile ordered state where the ligands are aggregated into crystalline bundles.\cite{Luedtke1998, Ghorai2007, Lane2010, Widmer-Cooper2014, Bolintineanu2014}

Recent work indicates that this ability of the ligand shell to change its structure can result in interactions between nanoparticles that deviate substantially from those predicted by theoretical approaches which assume a uniform ligand density around the particles.\cite{Khan2009self, Goubet2011forces, Sigman2004metal} Simulations have shown that the interaction between particles in dispersion can change rapidly from repulsive to attractive as the ligands order,\cite{Widmer-Cooper2014, Widmer-Cooper2016} while experiments indicate that the structure of nanoparticle agglomerates is affected by short-range interactions between the ligand shells.\cite{Gerstner2018} Recently, using a combination of experiment and computer simulations, we demonstrated that the agglomeration of smaller hexa\-decane\-thiol-coated gold nanoparticles is driven by the ordering of the ligand shell, while for larger particles the van der Waals (vdW) attraction between the cores becomes strong enough to drive agglomeration before the ligands order.\cite{Kister2018} This transition from shell- to core-dominated agglomeration was shown to result in a nonlinear change in the interparticle spacing, and should have other important consequences. 

Here, we systematically investigate the effect of the ligand length on \Revised{colloidal} stability for a range of particle sizes. Naively, one may expect the stability of inorganic nanoparticles with apolar ligands to increase as the ligand shell becomes thicker and keeps the cores further apart. We show that this is only true for large particles, and that there exists an inversion in the dependence of the agglomeration temperature on the ligand length. In particular, we consider here both Au and CdSe nanocrystals suspended in decane and coated with linear alkanethiols ranging in length from 12 to 18 carbons. \Revised{These two systems were chosen because they can be synthesized as size tunable spherical colloids with relatively narrow size distributions. In addition, the Hamaker constants differ considerably, allowing us to explore the role of van der Waals forces on the agglomeration temperature.} We find that, for larger Au particles, the agglomeration temperature \emph{decreases} with ligand length, as one would expect if agglomeration were driven by vdW attraction between the cores; for smaller Au and CdSe particles, however, the agglomeration temperature \emph{increases} with ligand length, because the agglomeration is now driven by attraction between the ligand shells as they order, and longer alkane chains order at higher temperatures. Further, we show that ligand length has a strong effect on particle spacing in the core-dominated regime but only a small effect in the shell-dominated regime.

\section{Results and Discussion}

First, consider \enquote{large} gold nanoparticles (AuNP) with core diameters above \SI{8}{\nano\meter} coated with alkanethiol ligands. Dispersions with \numlist{8.3;8.9}\SI{\pm 0.8}{\nano\meter} cores and dodecanethiol (\(SC_{12}\)), hexadecanethiol (\(SC_{16}\)), and octadecanethiol (\(SC_{18}\)) shells with a coverage of \Revised{\SI[mode=text]{5.5}{ligands \nano\meter^{-2}}} were analyzed at a concentration of \SI{2.5}{\milli\gram\per\milli\liter} by \textit{in situ} small angle X-ray scattering (SAXS). \Revised{When fully extended, these ligands are approximately \SI{1.78}{\nano\meter}, \SI{2.28}{\nano\meter}, and \SI{2.54}{\nano\meter} in length, respectively.} The experiments were designed to be fully comparable to the experiments on particles with \(SC_{16}\) shells on different cores reported previously\cite{Kister2018}; additional details can be found there and in the experimental section. Briefly, thoroughly purified AuNPs in decane were introduced into an X-ray beam and the scattering was recorded using a large 2D detector while changing the temperature in small steps. Below a certain temperature, the particles began to agglomerate, clearly indicated by the appearance of a peak in the structure factor \(S(q)\)\cite{Johnson1959x}. We define the agglomeration temperature, \(T_\mathrm{agglo}\), as the temperature at which \SI{20}{\percent} of the particles are agglomerated.

Cores larger than \SI{8}{\nano\meter} coated in \(SC_{16}\) ligands have been shown to attract each other sufficiently strongly that the particles agglomerate while the ligands are still disordered (the \enquote{core-dominated regime}).\cite{Kister2018} Because the vdW attraction between the cores increases as the particles approach, one would therefore expect a decreased thermal stability for shorter ligands. This is precisely what we observe in the current study: Figures \ref{Figure_1}a and \ref{Figure_1}b show that \SI{8.3}{\nano\meter} cores with \(SC_{12}\) ligands agglomerated roughly \SI{60}{\celsius} above those with \(SC_{16}\) and \(SC_{18}\) ligands. Increasing the particle size to \SI{8.9}{\nano\meter} further increased the vdW attraction between the cores and resulted in additional destabilization of particles coated in shorter ligands (see Figure \ref{Figure_1}b and Figure S1 in the Supporting Information). At this size, \(SC_{12}\)-stabilized particles were too attractive to be dispersed, and the \(SC_{16}\)-stabilized particles agglomerated at higher temperatures than the \(SC_{18}\)-stabilized particles. This is consistent with previous observations that exchanging for longer organic ligands facilitates the transfer of gold nanoparticles from water into organic solvents.\cite{Lista2014,Karg2011}.

Molecular dynamics simulations of \SI{8.3}{\nano\meter} core particles in explicit \textit{n}-decane confirmed that the ligand shells are mobile and disordered at the experimental agglomeration temperatures. Snapshots of the particles at \(T_\mathrm{agglo}\) are shown in Figure \ref{Figure_1}d, top row, where the solvent molecules have been hidden to reveal the structure of the ligand shell more clearly. A quantitative measure of ligand order is provided by the average dihedral angle of the alkane tails, which indicates the degree to which the ligands are extended in all-trans conformations (\ang{180}). Figure \ref{Figure_1}e and Figure \ref{Figure_1}d (bottom row) show that the ligands did not extend and start clustering together until well below \(T_\mathrm{agglo}\), where the experimental agglomeration temperatures have been indicated by large crossed symbols. Overall, simulations and experiments consistently indicate that when the agglomeration is core-dominated, shorter ligands result in particles which are less stable to agglomeration.

The experimental core surface spacings reported in Figure \ref{Figure_1}c indicate minimal overlap between the \(SC_{18}\) and \(SC_{16}\) ligand shells, with the spacings close to twice the thickness of the shell around an isolated particle (see Figure S2 in the Supporting Information). This is consistent with the uniform and mobile ligand shells observed in our simulations, which should result in entropic steric repulsion between the particles as their shells start to overlap and the conformational freedom of the ligands becomes restricted. \cite{Widmer-Cooper2014,Kister2018} 

We have calculated the core-core vdW interaction at these separations using both spherical and icosahedral models for the gold core (see Figure \ref{Figure_1}f). TEM images indicate that the cores have substantial faceting, with some triangular facets (typical of icosahedra) visible (see Figure S3 in the Supporting Information). Our calculations show that the core-core interaction is substantially stronger when such faceted particles are oriented face-to-face than would be estimated from a spherical model. For the \(SC_{18}\) and \(SC_{16}\) coated cores, the interaction is roughly \(-1 \mathrm{k_B} T_\mathrm{agglo}\) for regular icosahedra, not far from the \(-1.5 \mathrm{k_B} T\) necessary to drive agglomeration in the limit of low particle concentration. In contrast, the experimental spacings indicate substantial compression of the \(SC_{12}\) ligands, with the spacing now substantially less than twice the thickness of the shell around an isolated particle (roughly \SI{2.6}{\nano\meter} versus \SI{3.2}{\nano\meter}). This compression appears to be the result of stronger core-core attraction once the shells are thinner, with analytical calculations indicating a core-core vdW interaction around \(-2.2 \mathrm{k_B} T_\mathrm{agglo}\) at the experimental spacing. Decreasing the ligand length in the core-dominated regime thus appears to have the same effect as increasing the particle size \cite{Kister2018}, \textit{i.e.}, the vdW attraction between the metal cores becomes stronger, which eventually results in substantial compression of the disordered ligand shell. 

\afterpage{\begin{figure}[H]
\centering
    \includegraphics[width=0.8\linewidth]{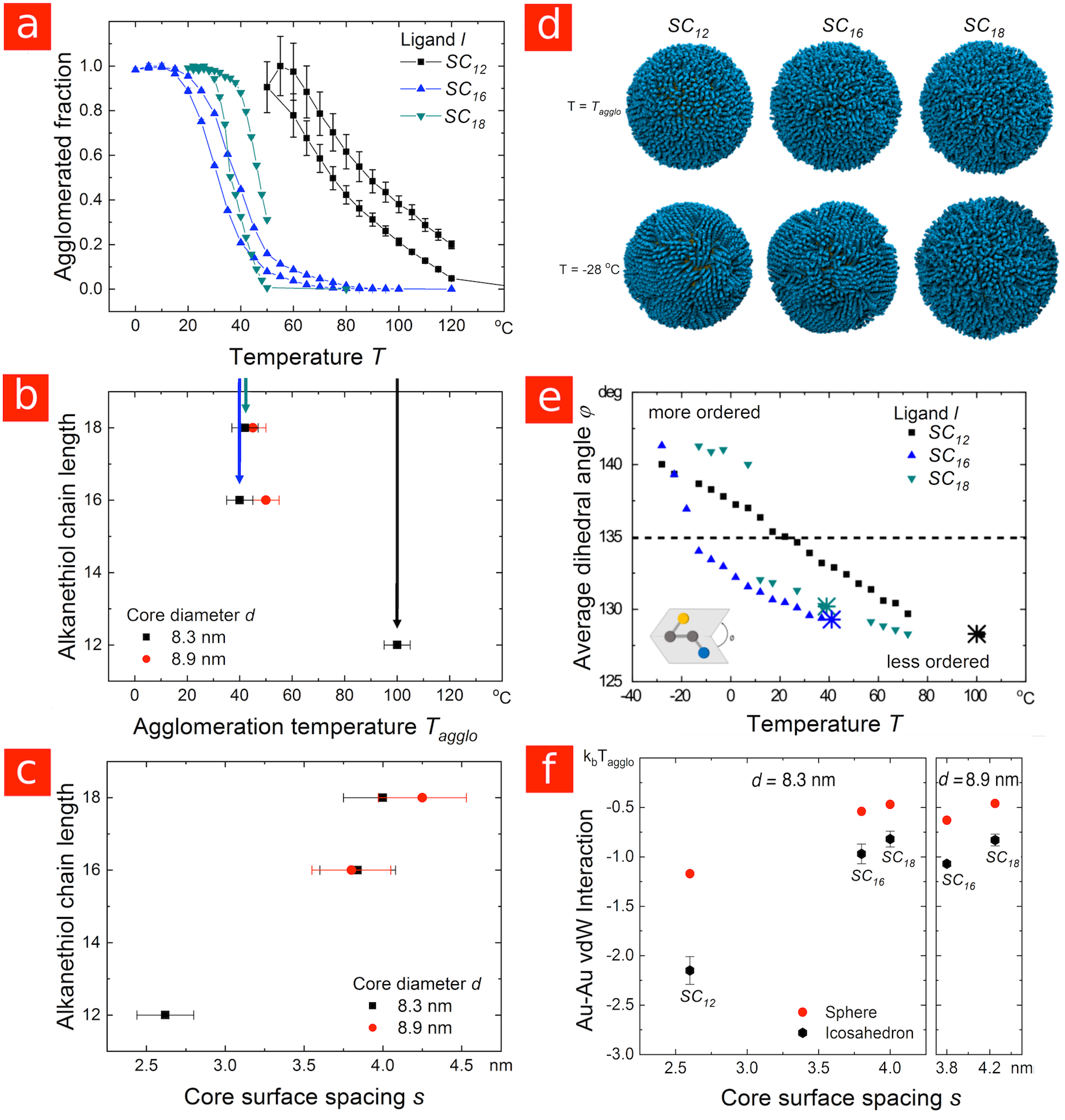}
    \caption{Temperature-dependent agglomeration of AuNP with \numlist{8.3;8.9} \si{\nano\meter} diameter cores studied by SAXS (left row) and theoretical models (right row). (a) Fraction of agglomerated \SI{8.3}{\nano\meter} particles covered with alkanethiol ligands of different lengths in decane as determined by \textit{in situ} SAXS. All particles were dispersed at high temperatures and agglomeration occurred upon cooling, as indicated by the increase in structure factor. (b) Agglomeration temperature (where \SI{20}{\percent} of the particles were agglomerated) as a function of ligand chain length. (c) Core surface spacing between AuNPs at the agglomeration temperature. The much smaller separation between \(SC_{12}\) coated particles indicates substantial compression of the ligand shell. \Revised{The error bars in (b) and (c) represent one standard error.} (d) Molecular Dynamics simulation snapshots at \(T_\mathrm{agglo}\) (top row) and \SI{-28}{\celsius} (bottom row), and (e) the average dihedral angle of the ligands, demonstrate that, regardless of the ligand length, \SI{8.3}{\nano\meter} AuNP agglomerate before the ligands order (the experimental agglomeration temperatures are indicated by large crossed symbols). The scheme at the bottom left shows the definition of the dihedral angle $\phi$ and the dashed line is a guide to the eye. In the snapshots, decane solvent is hidden for clarity. (f) Analytically estimated core-core vdW interaction at the experimental particle spacings for \numlist{8.3;8.9} \si{\nano\meter} AuNPs. The values are substantially higher for icosahedral cores oriented face-to-face than for spherical cores. \Revised{Error bars for the icosahedral case represent confidence intervals (see Methods for details).}}
    \label{Figure_1}
    \vspace{0.5em}
\end{figure}}

Next, consider \enquote{smaller} gold particles with core diameters of around \SI{6}{\nano\meter}. When coated with \(SC_{16}\) ligands, their agglomeration in decane is dominated by the attraction that arises between the shells as the ligands cluster together to form ordered bundles like those shown in Figure \ref{Figure_2}a (the \enquote{shell-dominated regime})\cite{Kister2018}. Our molecular dynamics simulations show that the temperature at which the ligands order depends strongly on ligand length: longer ligands order at higher temperatures for the same \SI{5.8}{\nano\meter} core diameter, as shown in Figure \ref{Figure_2}b. This is consistent with previous studies of ligand behavior in \Revised{the absence of solvent},\cite{Luedtke1998,Ghorai2007,Lane2010,Bolintineanu2014} and reflects a change in the balance between chain energy and entropy that is also seen in the tendency of longer alkanes to freeze at higher temperatures. Consequently, our simulations predict that smaller particles coated in longer ligands will agglomerate at higher temperatures, the exact opposite of the trend found for larger particles. 

Experiments with \SI{6}{\nano\meter} AuNP in heptane are consistent with these predictions.\cite{Born2013} In particular, the experimental agglomeration temperatures, indicated by the crossed symbols in Figure \ref{Figure_2}b, increased with ligand length in a way that correlates with the increase in the ligand ordering temperature. Moreover, our simulations indicated that the ligands are ordered at the experimental agglomeration temperatures in all cases (see Figure \ref{Figure_2}a). Note that the experimental results in this case are for particles in heptane rather than decane, which probably offsets the small difference in core diameter.

\afterpage{\begin{figure}[H]
\centering
    \includegraphics[width=0.6\linewidth]{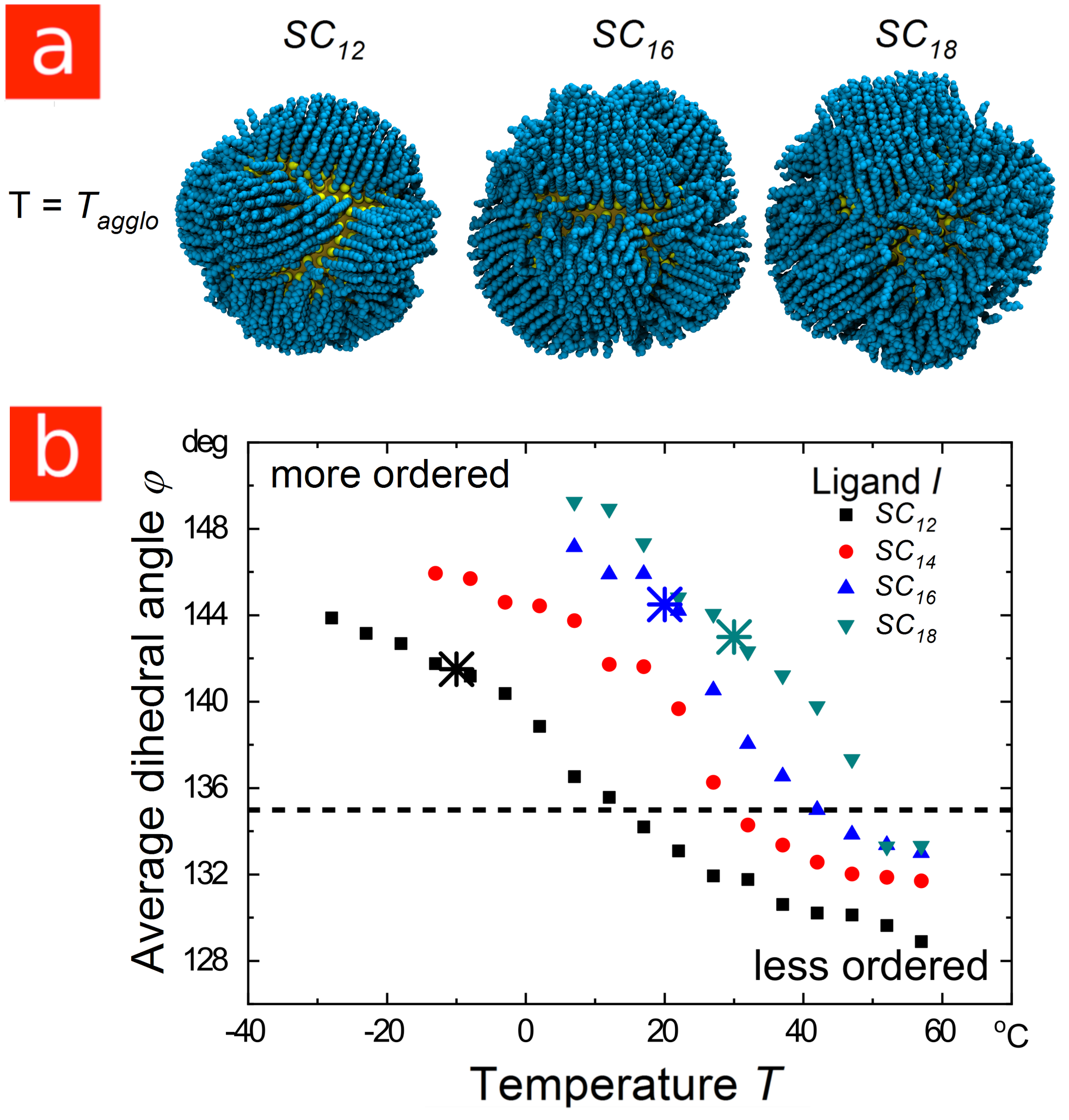}
    \caption{(a) Simulation snapshots of \SI{5.8}{\nano\meter} AuNPs in decane show that the ligands are ordered at the experimental agglomeration temperature. (b) The degree of ordering at different temperatures as quantified by the average dihedral angle of the ligands. \(T_\mathrm{agglo}\), indicated by the large crossed symbols, corresponds to the experimental agglomeration temperature for \SI{6}{\nano\meter} AuNPs dispersed in heptane\cite{Born2013}.}
    \label{Figure_2}
    \vspace{0.5em}
\end{figure}}

Based on our results, we propose the following rule for the ligand-dependent \Revised{colloidal} stability of apolar nanoparticles covered in linear alkyl ligands: \enquote{Long ligands stabilize larger core-dominated particles and destabilize smaller shell-dominated particles.} This rule is not restricted to AuNPs. As we show in Figure \ref{Figure_3}, the agglomeration of cadmium selenide particles (CdSeNPs) with \SI{5.8}{\nano\meter} diameter cores in decane is shell-dominated, with a temperature-dependent stability that follows the same trends as shell-dominated AuNPs. Experimentally, the agglomeration temperatures increased with ligand length (see Figures \ref{Figure_3}a and \ref{Figure_3}b), consistent with the trend in ligand ordering temperatures predicted by simulation (see Figures \ref{Figure_3}d and \ref{Figure_3}e). Moreover, the simulations indicate that the ligands order prior to particle agglomeration.

\afterpage{\begin{figure}[H]
\centering
    \includegraphics[width=0.8\linewidth]{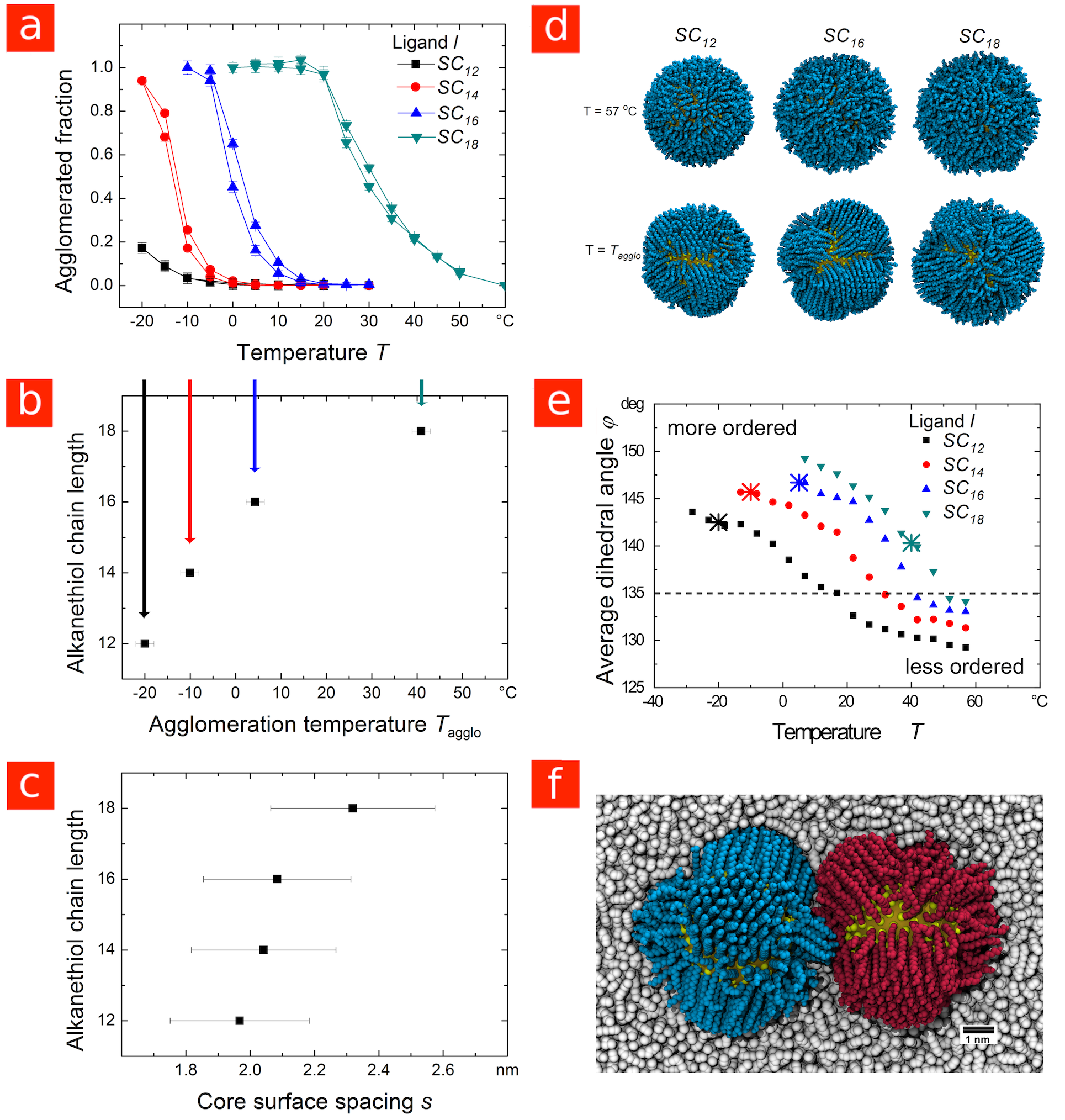}
    \caption{(a) Fraction of \SI{5.8}{\nano\meter} CdSeNPs agglomerated in decane as a function of temperature and ligand length, as determined by \textit{in situ} Small-Angle X-ray Scattering. (b) Agglomeration temperature as a function of ligand chain length. As for the small AuNP, there is an increase in \(T_\mathrm{agglo}\) with ligand length. (c) Core surface spacing between particles at the agglomeration temperature. \Revised{The error bars in (b) and (c) represent one standard error.} (d) Simulation snapshots at \SI{57}{\celsius} (top row) and at \(T_\mathrm{agglo}\) (bottom row). (e) Average dihedral angle of the ligands. Comparison with the experimental agglomeration temperatures, indicated by large crossed symbols, show that the particles agglomerate after the ligands order. (f) Simulation snapshot at \(T_\mathrm{agglo} + \SI{10}{\celsius}\) showing two \(SC_{16}\) covered CdSeNPs interacting at \SI{2.1}{\nano\meter} separation. Different colors have been used for the ligands on the two particles, and the solvent was partially hidden.}
    \label{Figure_3}
    \vspace{0.5em}
\end{figure}}

The experimental CdSeNP spacing is consistent with the ligands ordering prior to agglomeration, with a mean value close to, and in most cases smaller than, the length of one extended ligand (Figure \ref{Figure_3}c). At the coverage that we measured experimentally, this close spacing suggests interdigitation of ligand bundles as shown in Figure \ref{Figure_3}f. Previously, we have shown that such configurations correspond to the minimum in the potential of mean force acting between AuNPs with similar size and ligand coverage.\cite{Kister2018}. Interdigitation of alkanethiol chains was inferred from HRTEM images of dried agglomerates in 1993;\cite{Giersig1993} the results here indicate that in the liquid it occurs at the bundle level and that the extent depends on particle size.

Recently, an alternative explanation has been offered for particle spacings close to one ligand length in assemblies of \SIrange{2}{5}{\nano\meter} CdSe particles coated with alkylamine ligands.\cite{Geva2017} Simulations in \Revised{the absence of solvent} indicated ligand shell collapse, with the majority of ligands wrapping around the core rather than extending out from it. While this type of ligand collapse may be possible on small particles at low surface coverage in poor solvents, our simulations in decane do not show such behavior. Even at only 70\% of our measured coverage, the ligands form distinct ordered bundles that tilt with respect to the particle surface, but do not wrap around it (see Figures S7 in the Supporting Information, \ref{Figure_3}d, and \ref{Figure_3}f), resulting in an average ligand shell thickness considerably more than half a ligand length (Figure S8 in the Supporting Information).

Thus, while we do not expect drastic changes in the ligand shell geometry for the range of ligand shell densities encountered here, we note that even small differences in the ligand coverage affect agglomeration temperatures. Our experimental results show that the agglomeration of CdSeNPs is sensitive to the ligand concentration in solution (see Figure S6 in the Supporting Information), and our simulations show that the ligands order at lower temperature when the surface coverage is reduced from \Revised{\SI[mode=text]{5.5}{ligands \nano\meter^{-2}}} to \Revised{\SI[mode=text]{3.6}{ligands \nano\meter^{-2}}} (compare Figures \ref{Figure_3}e and S7). These simulations are consistent with earlier results for CdS nanorods,\cite{Widmer-Cooper2016} and indicate a potential link between ligand concentration in solution and particle stability \Revised{to agglomeration}.

The agglomeration of much larger silica particles also appears to be shell-dominated. The temperature-dependent stability of \SI{72}{\nano\meter} and \SI{246}{\nano\meter} diameter silica particles with similar (stearyl alcohol with 18 carbon atoms) ligands was studied previously by van Blaaderen and Bonn, who used non-linear optical spectrometry to identify the order-disorder ligand transition.\cite{Roke2006} They detected agglomeration (\enquote{gelation}) via increased optical scattering at a temperature close to the molecular transition. Using a spherical model, we estimate that the attractive vdW interaction between silica cores in a general hydrocarbon medium will not be strong enough to induce room temperature agglomeration until a diameter of around \SI{500}{\nano\meter} if the particles are coated with \(SC_{18}\) ligands. This is due to the low Hamaker constant of amorphous silica (\SI{0.41}{\electronvolt}), which is not much higher than that of typical organic solvents. Core materials with similarly low Hamaker constants are likely to exhibit the shell-dominated behaviour we have described above for different ligand lengths, even at very large diameters. \Revised{On the other hand, CdSe particles are likely to exhibit a crossover to core-dominated agglomeration at a similar diameter to Au particles (i.e. around \SI{8}{\nano\meter}) due to their significant dipole moments.}

Conventional colloid theory, as represented by Khan \latin{et~al.},\cite{Khan2009self} predicts the right trends for the agglomeration temperature and interparticle spacing in the core-dominated regime, but fails completely in the shell-dominated regime (see Figure \ref{Figure_4}). Even for the core-dominated case, the predicted particle separations are larger than those observed experimentally. We conclude that improved models will need to account for temperature-dependent transitions in the ligand shell and provide a more accurate description of the core shape and the compressibility of the ligand shell in different states.

\afterpage{\begin{figure}[H]
\centering
    \includegraphics[width=0.6\linewidth]{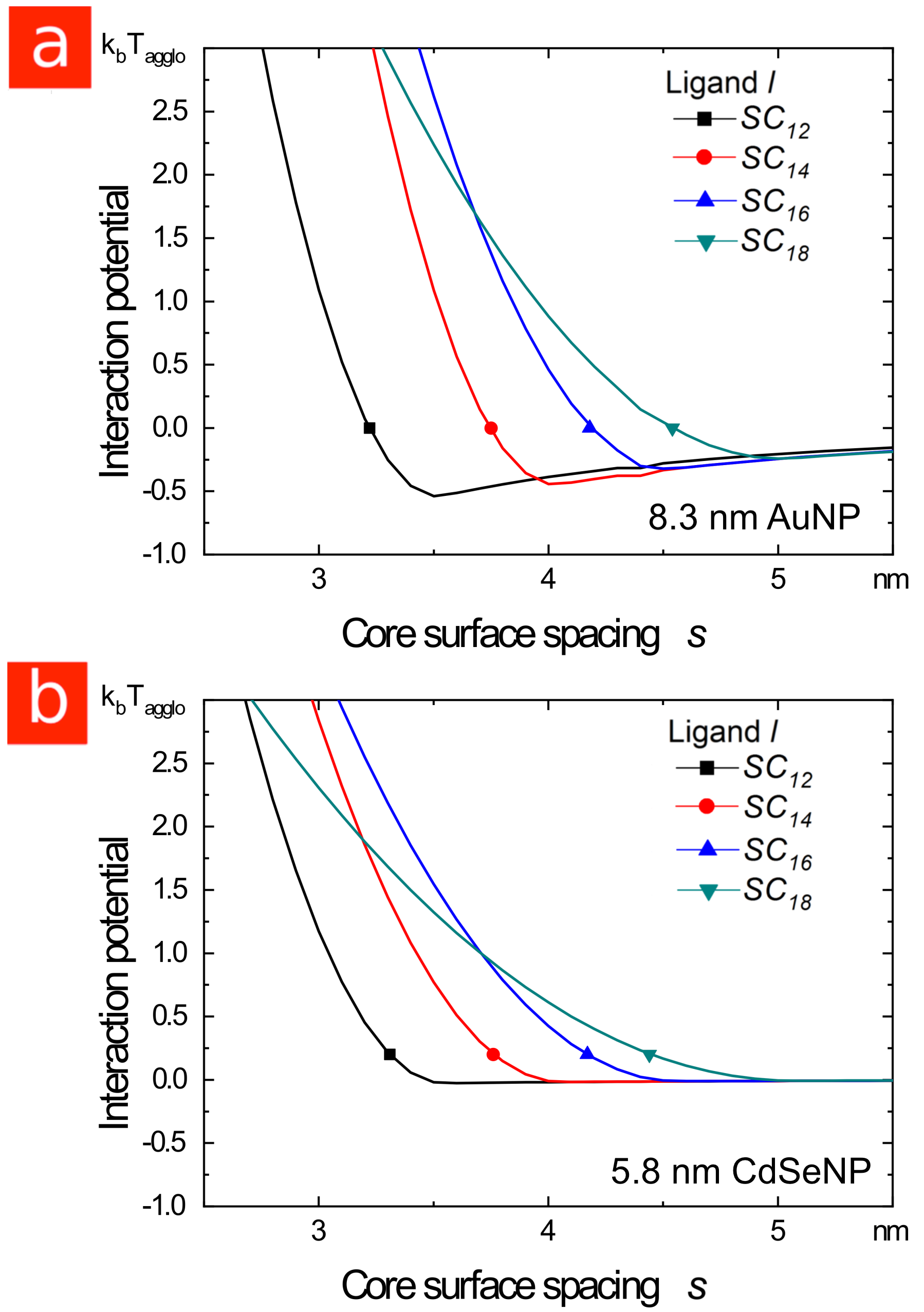}
    \caption{Interaction potentials for (a) \SI{8.3}{\nano\meter} AuNPs and (b) \SI{5.8}{\nano\meter} CdSeNPs calculated using conventional colloid theory, as represented by Khan et al.\cite{Khan2009self}. The predicted potentials are incompatible with our experiments and simulations.}
    \label{Figure_4}
    \vspace{0.5em}
\end{figure}}

Our results also differ in important ways from simulation studies of particles interacting without a solvent. In \Revised{the absence of solvent}, the interaction between the particles is strongly attractive irrespective of whether the ligands are ordered or disordered,\cite{Widmer-Cooper2014, Kister2018} and chain length seems to have a negligible effect on the particle spacing for gold cores ranging from \SIrange{2}{9}{\nano\meter} in diameter and ligands lengths ranging from 4 to 12 or 9 to 19 carbons.\cite{Vlugt2008, Jabes2014, Waltmann2017, Waltmann2018} In contrast, we find that in good solvents both the particle interaction and spacing depend strongly on the ligand length, with very different effects depending on particle size.

\section{Conclusions}
The effect of ligand length on the colloidal stability of apolar metallic and semiconducting nanoparticles depends strongly on core size. Our experiments and simulations consistently showed that the relationship inverts when the agglomeration changes from core- to shell-dominated. In the core-dominated regime, \textit{i.e.} for gold cores with diameters larger than approximately \SI{8}{\nano\meter}, increasing the ligand length increased the stability of the suspension by extending the range of repulsion between the disordered ligand shells, thereby reducing the vdW attraction between the cores. In contrast, the \Revised{colloidal} stability of particles with smaller cores in the shell-dominated regime increased for shorter ligands that order at lower temperatures. Classical theories do not account for the possibility of the ligand shell structure becoming anisotropic, or its sensitivity to changes in temperature and solvent nature, and therefore fail to predict \Revised{colloidal} stability.

Our results also indicate that it is insufficient to simply consider the core/shell ratio when seeking to understand the effect of the ligand length on particle stability \Revised{to agglomeration}. Increasing the core size always decreases \Revised{colloidal} stability, whereas increasing the ligand length can either increase or decrease \Revised{colloidal} stability depending on whether the particle is in the core- or shell-dominated regime.

More generally, our results show that even small changes in the ligand length can significantly affect the free energy balance between the disordered and ordered ligand states. This is especially relevant for particles that fall in the shell-dominated regime. We expect that other changes to the ligand structure, such as the presence of double bonds or branches, will also strongly affect particle stability \Revised{to agglomeration} by modifying the relative free energies of the ordered and disordered ligand states. This may help explain other intriguing results in the recent literature,\cite{Yang2016, YangQin2016} and further studies are currently underway.

\section{Methods}

All chemicals were obtained from Sigma Aldrich (unless noted otherwise) and used without further purification. Note that all methods were chosen to provides samples that are as comparable as possible to the data published previously \cite{Kister2018}.

\subsection{Nanoparticle synthesis} 

Gold nanoparticles (AuNP) with core diameters of \SI{8.3}{\nano\meter} and \SI{8.9}{\nano\meter} were synthesized as before using a modified protocol based on the method of Wu and Zheng \cite{Zheng2013}. To prepare gold cores with a diameter of \SI{8.3}{\nano\meter}, a mixture of \SI{8}{\milli\liter} benzene (puriss. $\ge$ 99.7\%), \SI{8}{\milli\liter} oleylamine (technical grade, 70\%), and \SI{100}{\milli\gram} of HAuCl4 (with crystal water) was stirred at \SI{20}{\celsius} and \SI{500}{\radian\per\minute} for \SI{1}{\minute} under argon atmosphere. A solution of \SI{40}{\milli\gram} tert-butylamine borane (ABCR, 97\%) in \SI{2}{\milli\liter} benzene and \SI{2}{\milli\liter} oleylamine (OAm) was then added. The color of the solution immediately became dark purple. After stirring for \SI{60}{\minute} at \SI{20}{\celsius}, the nanoparticles were purified once by precipitating with \SI{30}{\milli\liter} ethanol and centrifugation at \SI{4000}{\radian\per\minute} for \SI{5}{\minute}. The precipitated nanoparticles were then redispersed in \SI{20}{\milli\liter} heptane (puriss. $\ge$ 99\%). Gold cores with a diameter of \SI{8.9}{\nano\meter} were produced by a second overgrowth step. \SI{60}{\milli\gram} of HAuCl4, \SI{5}{\milli\liter} of benzene, and \SI{1}{\milli\liter} of oleylamine were added to \SI{10}{\milli\liter} of the \SI{8.3}{\nano\meter} dispersion in heptane and stirred for \SI{5}{\hour} at \SI{60}{\celsius}. The resulting dispersion was then purified as above.

Cadmium selenide nanoparticles (CdSeNPs) with core diameters of \SI{6}{\nano\meter} were synthesized as follows. First, three stock solutions were prepared, a Se injection solution (i), a Cd growth solution (ii), and a Se growth solution (iii): 
(i) \SI{0.3265}{\gram} Se were dissolved in a mixture of \SI{2.5}{\gram} trioctylphosphine, \SI{2.5}{\gram} octadecene, and \SI{6}{\gram} oleylamine in a nitrogen-filled glovebox to give a clear, slightly yellow solution. (ii) A solution containing \SI{0.17}{\mol\per\liter} of cadmium were made from \SI{0.22}{\gram} cadmium oxide, \SI{0.97}{\gram} oleic acid, and \SI{6.23}{\gram} 1-octadecene in a 3 neck round bottom flask on a Schlenk line. The solution was degassed under vacuum (\(<\) \SI{1}{\milli\bar}) for \SI{60}{\minute} at \SI{80}{\celsius}, heated to \SI{250}{\celsius} and held until clear, then cooled to room temperature. Whilst cooling \SI{1.13}{\milli\liter} of oleylamine were added. The final solution was clear and slightly yellow. (iii) A solution containing \SI{1.7}{\mol\per\liter} of selenium was prepared by dissolving \SI{0.25}{\gram} of selenium in \SI{1.55}{\gram} trioctylphosphine in a nitrogen-filled glovebox to give a clear colourless solution.

The synthesis started with \SI{0.22}{\gram} cadmium oxide, \SI{3}{\gram} oleic acid, and \SI{30}{\gram} octadecene in a 3 neck round bottom flask that was degassed under vacuum (\(<\) \SI{1}{\milli\bar}) for \SI{60}{\minute} at \SI{80}{\celsius}. The mixture was then heated to \SI{260}{\celsius} until a clear solution (iv) had formed. The selenium injection solution (i) was loaded into a \SI{24}{\milli\liter} disposable syringe equipped with a 16 G needle and rapidly injected into the cadmium solution (iv) at \SI{260}{\celsius}. The temperature of the reaction solution was allowed to recover to \SI{250}{\celsius} where it was held for NP growth. After \SI{20}{\minute}, \SI{2}{\milli\liter} of \SI{0.17}{\mol\per\liter} cadmium growth stock (ii) and \SI{0.2}{\milli\liter} of \SI{1.7}{\mol\per\liter} selenium growth stock (iii) were added dropwise to the reaction. The addition of cadmium and selenium growth solution (ii, iii) was continued every \SI{10}{\minute}. After 3 additions, the reaction was left for a further \SI{10}{\minute} at \SI{250}{\celsius}, then cooled to room temperature. The NPs were washed three times via precipitation with acetone and \Revised{resuspended} in toluene.

\subsection{Nanoparticle characterization}

Small Angle X-ray Scattering (Xenocs Xeuss 2.0) and Transmission electron microscopy (JEOL JEM 2010) were used to measure the core size of the NPs as previously described\cite{Kister2018}. Scattering data from SAXS was analyzed using SASfit (Version 0.94.6, Paul Scherrer Institute) and TEM micrographs were analyzed using ImageJ distributed by NIH (Version 1.45s)

\begin{table}[ht]
\centering
\small
  \caption{AuNPs used for this study, with diameters obtained from transmission electron microscopy and small angle X-ray scattering.}
  \label{table_size}
  \begin{tabular*}{0.8\textwidth}{@{\extracolsep{\fill}}cccc}
    \hline
    Number & d (TEM) & d (SAXS)\\
    \hline
    Au01 & 8.5 nm $\pm$ \SI{7.1}{\percent} & 8.3 nm $\pm$ \SI{6.7}{\percent}\\
    Au02 & 8.9 nm $\pm$ \SI{8.5}{\percent} & 8.9 nm $\pm$ \SI{6.8}{\percent}\\
    CdSe & 5.8 nm $\pm$ \SI{7.1}{\percent} & 6.0 nm $\pm$ \SI{9.6}{\percent}\\
    \hline
  \end{tabular*}
\end{table}

\subsection{Ligand exchange}

\textbf{AuNPs.} Ligand exchange on AuNPs was performed as described previously \cite{Kister2016pressure}. AuNPs coated with oleylamine were heated to \SI{80}{\celsius} and an excess of required alkanethiol was added. After stirring for further \SI{10}{\minute}, the particles were purified and redispersed in decane ($\ge$ 95\%).

\textbf{CdSeNPs.} As-synthesized CdSeNPs were precipitated with acetone/ethanol and resuspended in a solution of the respective alkanethiol ligand (40 wt-\(\%\) in chloroform) with triethylamine (1 molar equivalent with respect to thiol). The resulting NP dispersion was heated for 3 hours at \SI{45}{\celsius} while stirring. The NPs were then washed via precipitation with antisolvent and centrifugation (3,300 x g for \SI{3}{\minute}). The antisolvent was chosen to optimally dissolve excess ligand: 1:1 (v/v) methanol/ethanol mixture for hexanethiol and octanethiol ligands, or 1:1 (v/v) acetone/ethanol mixture for dodecanethiol and longer ligands. The NPs were resuspended again in a solution of ligand (40 wt-\(\%\) in chloroform), stirred at \SI{45}{\celsius} for 2 hours, then washed as before and resuspended in a 0.1 M solution of ligand in chloroform. After stirring at room temperature for 24-48 hours the NPs were washed three times and resuspended in pure chloroform. \Revised{Chambrier et al. have shown this procedure leads to almost complete displacement (>92\%) of amines by the alkane thiol ligands.\cite{Chambrier2015}.}

\subsection{Thermogravimetric analyses}

Thermogravimetric analyses were performed using a Netzsch STA 449 F3 Jupiter. The measurements started at room temperature and run until \SI{800}{\celsius}. The heating rate was kept at \SI{10}{\kelvin \per \minute}. All measurements were done under an inert atmosphere. \Revised{Figure S4 shows representative TGA data of 1-hexadecanethiol coated AuNP with a core diameter of \SI{8.9}{\nano\meter}}.

\subsection{Small-Angle X-ray Scattering}

Experiments were performed using a Xeuss 2.0 from Xenocs SA (Grenoble, France) equipped with a copper $K_\alpha$ X-ray source and a PILATUS 1M detector from DECTRIS (Baden, Switzerland). 

To prevent solvent evaporation during the measurements, the samples (usually a quantity of \SI{20}{\micro \litre} to \SI{40}{\micro\litre}) were filled into glass capillaries (diameter of \SI{2}{\milli\meter}), which were then sealed with epoxy. 

For each measurement, the samples were introduced into a temperature controlled sample holder (Omega CN8200), Peltier-controlled with a temperature range between \SI{-20}{\celsius} and \SI{120}{\celsius}. The measurements started at high temperature to ensure a fully deagglomerated state. Afterwards the temperature was first decreased and later increased in \SI{5}{\celsius} steps. At each step, the samples were first equilibrated (\SI{20}{\minute}) followed by an acquisition (\SI{10}{\minute}). \Revised{Experiments with time-dependant observation of the agglomeration process as shown in figure S5 were used to ensure that most particles that had lost colloidal stability at this temperature had agglomerated before the next temperature step was taken.} Data treatment was carried out as described previously \cite{Schnablegger2013saxs, Kister2018}.

\subsection{Molecular dynamics simulations}
The nanoparticles were modeled as spherical cores covered in alkanethiol ligands of various lengths ($-S(C_nH_{2n+1})$, n = 12, 14, 16, 18) and immersed in explicit \textit{n}-decane solvent. Both \SI{5.8}{\nano\meter} and \SI{8.3}{\nano\meter} Au cores and \SI{5.8}{\nano\meter} CdSe were considered, with the ligands assumed to be irreversibly bound to the Au and CdSe cores. According to TGA measurements, the surface coverages for Au and CdSe cores are very similar: \Revised{\SI[mode=text]{5.5}{ligands \nano\meter^{-2}}} and \Revised{\SI[mode=text]{5.2}{ligands \nano\meter^{-2}}}, respectively. We therefore used a coverage of \Revised{\SI[mode=text]{5.5}{ligands \nano\meter^{-2}}} in all cases. To assess the effect of changing the ligand coverage on the structure and transition of the ligand shell, simulations were also performed for \SI{5.8}{\nano\meter} CdSeNPs with \(SC_{18}\) ligands at a coverage of \SI{3.6}{\nano\meter^{-2}}. The positions of the sulfur atoms were determined by placing them on a spherical shell around the implicit core (\SI{0.15}{\nano\meter} further out). They were then allowed to find their optimal positions on this shell, subject to a repulsive interaction (standard Coulombic potential with a relative dielectric constant \Revised{\(\epsilon = 10\)}, truncated at \SI{24}{\angstrom}), which ensured that the binding sites were approximately equidistant from one another. The sulfur atoms were subsequently treated as part of the rigid core of the particle, using the RATTLE algorithm\cite{Andersen1983} to constrain their positions. \Revised{This simplification ignores the possibility of ligand detachment, but should be reasonable given the high surface coverages  considered in this work.} The rest of the ligand and solvent molecules were modeled using a united-atom representation, with each CH$_x$ group being represented by a single particle. These particles interacted with one another according to the 12-6 Lennard-Jones (LJ) potential, with parameters as used and described previously\cite{Widmer-Cooper2014}. Bond stretching, bond bending, and dihedral torsion terms were also considered within each molecule\cite{Martin1998}. The interaction between the CH$_x$ groups and the cores was efficiently modelled using a 9-3 LJ potential, using the parameters in Table \ref{core_interactions}. 

\begin{table}[ht]
\centering
\small
  \caption{Lennard-Jones parameters used to describe non-bonded interactions between Au/CdSe cores and CH$_x$ particles in our simulations, according to the truncated pairwise potential $V\left(r_{ij}\right)=\epsilon\left[\frac{2}{15}\left(\frac{\sigma_{ij}}{r_{ij}}\right)^9-\left(\frac{\sigma_{ij}}{r_{ij}}\right)^3\right]$, for $r_{ij}<r_c$. In this equation, r$_{ij}$ is the distance between particles i and j, and $r_c=\SI{30}{\angstrom}$ is the cutoff distance.}
  \label{core_interactions}
  \begin{tabular*}{0.8\textwidth}{@{\extracolsep{\fill}}cccc}
    \hline
    Core material & $\epsilon$/k$_b$ (\si{\kelvin}) & $\sigma$ (\si{\angstrom}) & ref \\
    \hline
    Au & 88 & 3.54 & \cite{Pool2007} \\
    CdSe & 56 & 3.54 & \cite{Widmer-Cooper2014} \\
    \hline
  \end{tabular*}
\end{table}

Systems containing approximately 150,000 (\SI{5.8}{\nano\meter} cores) to 350,000 (\SI{8.3}{\nano\meter} cores) united atoms were investigated using molecular dynamics (MD) simulations with periodic boundary conditions. All simulations were performed using the LAMMPS simulation package \cite{Plimpton1995}. Individual NPs in explicit decane were initially equilibrated at constant volume at a temperature sufficiently high to ensure that the ligands were in the disordered state (e.g., \SI{400}{\kelvin}). During this run, the periodic simulation cell was slowly compressed until the solvent density far from the NP core was equal to the experimental density of pure decane at the corresponding temperature. Subsequent constant temperature runs were performed at a pressure of \SI[mode=text]{80}{atm}, using a Nos\'e-Hoover thermostat and barostat, which yielded bulk solvent densities within \SI{1}{\percent} of experimental values. The particles were first equilibrated for \SIrange{9}{14}{\nano\second} at temperatures ranging from \SIrange{245}{330}{\kelvin}, before \SI{1}{\nano\second} production runs were performed. Molecular graphics were produced using Visual Molecular Dynamics (VMD).\cite{Humphrey1996}

\subsection{Calculation of core-core interaction}
The vdW interaction between pairs of \SI{8.3}{\nano\meter} and \SI{8.9}{\nano\meter} gold cores in a hydrocarbon medium was estimated at the experimental particle separations using Hamaker-Lifshitz theory. Both spherical and icosahedral shapes were considered for the cores. For the spherical case, the interaction was calculated analytically using equation \ref{equation:vdW}, where \(A\) is the reduced Hamaker coefficient (we used a value of \SI{2}{\electronvolt}) and \(\tilde{s}\) is the rescaled spacing between the interacting particles (center-to-center distance divided by their core diameter). For the icosahedral case, the interaction was calculated using a coarse-grained atomistic representation, with 4087 interaction sites per particle \Revised{and all interaction pairs included in the summation}. Only the face-to-face relative orientation was considered, which should give the strongest interaction. \Revised{Confidence intervals were estimated by placing the spherical interaction sites either completely inside or centered on the surface of the icosahedra and extrapolating to a fully atomistic representation.} 
 
\begin{equation}
     U_\mathrm{vdW}= -\frac{A}{12} \left( \frac{1}{\tilde{s}^2-1}+\frac{1}{\tilde{s}^2}+2 \ln \left( 1-\frac{1}{\tilde{s}^2}\right) \right)
     \label{equation:vdW}
\end{equation}

\acknowledgement
P.M., N.K., D.M. and A.W. were supported by the ARC Centre of Excellence in Exciton Science (CE170100026). A.W. thanks the Australian Research Council for a Future Fellowship (FT140101061), and D.M. thanks the University of Sydney Nano Institute for a Postgraduate Top-Up Scholarship. Computational resources were generously provided by the University of Sydney HPC service, the National Computational Infrastructure National Facility (NCI-NF) Flagship program, and the Pawsey Supercomputer Centre Energy and Resources Merit Allocation Scheme. T.K. and T.K. thank the DFG Deutsche Forschungsgemeinschaft for funding and E. Arzt for his continuing support of this project. P.M. and T.K. also thank the DAAD for travel support.

\section*{Supporting Information Available}
Supporting Information shows the fraction of agglomerated \SI{8.9}{\nano\meter} AuNP, the radial density distributions of ligand and solvent around \SI{8.3}{\nano\meter} AuNP (from simulation), TEM data of the oleylamine capped AuNP and CdSeNP, position of the first peak of the structure factor for CdSeNP systems as a function of ligand concentration, simulation snapshots, average dihedral angles and radial density distributions of ligand and solvent for CdSeNPs with low ligand coverage (from simulation). This material is available free of charge via the Internet at http://pubs.acs.org/.

\bibliography{bib}

\end{document}

\begin{figure}[H]
    \centering
    \includegraphics[width=0.8\linewidth]{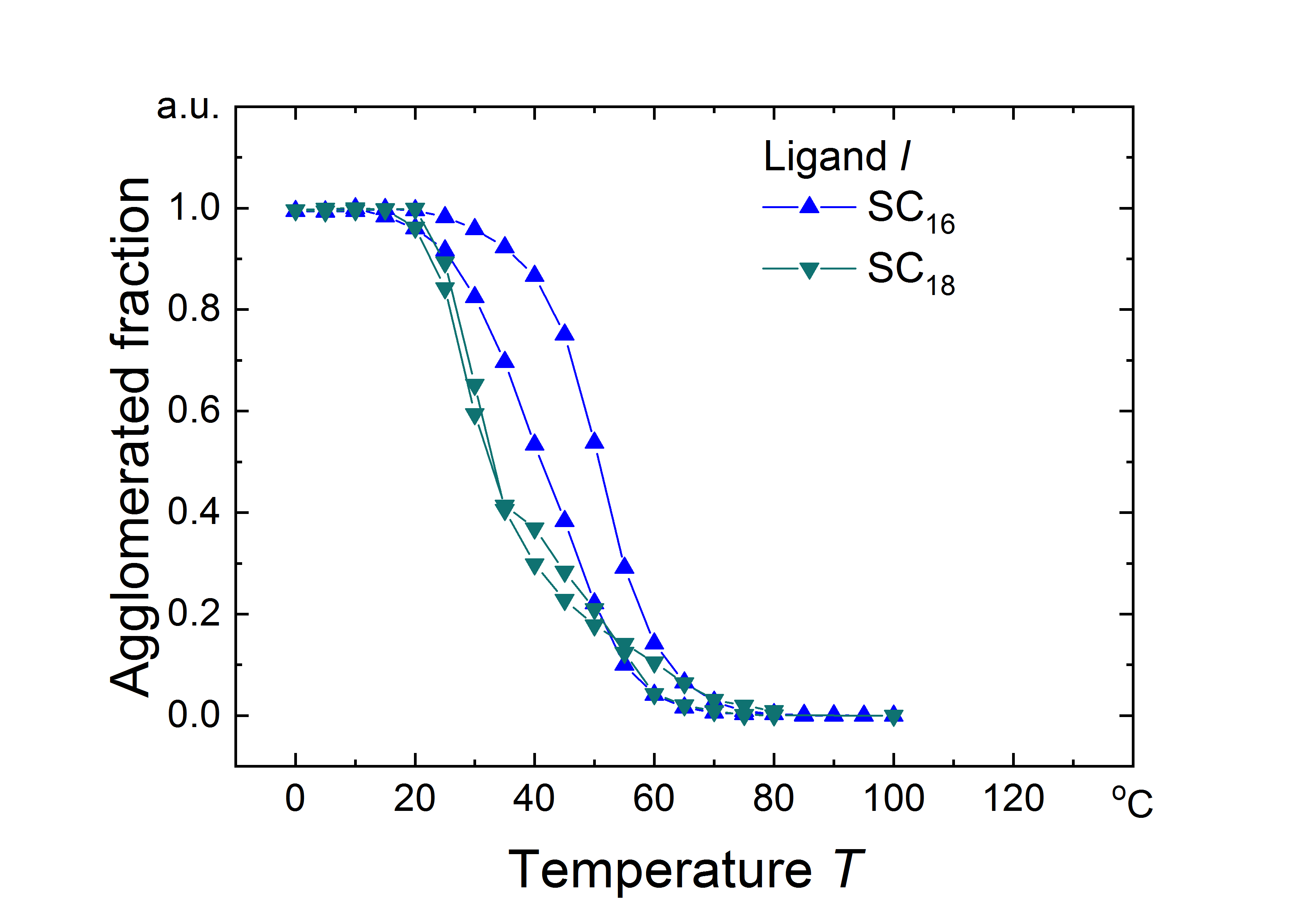}
    \caption{Fraction of agglomerated \SI{8.9}{\nano\meter} particles covered with alkanethiol ligands of different lengths in decane as determined by \textit{in situ} Small-Angle X-ray Scattering.}
    \label{SI_89nmAuNP}
    \vspace{0.5em}
\end{figure}

\newpage
Figure \ref{SI_89nmAuNP} shows how the agglomeration of \SI{8.9}{\nano\meter} core diameter gold particles covered with \(SC_{16}\) and \(SC_{18}\) ligands depends on temperature. Particles stabilized with \(SC_{12}\) ligands showed strong agglomeration, with a core surface spacing of \SI{2.72}{} $\pm$ \SI{0.7}{\nano\meter}, but could not be dispersed even with temperatures of \SI{120}{\celsius}.

\begin{figure}[H]
\centering
    \includegraphics[width=0.4\linewidth]{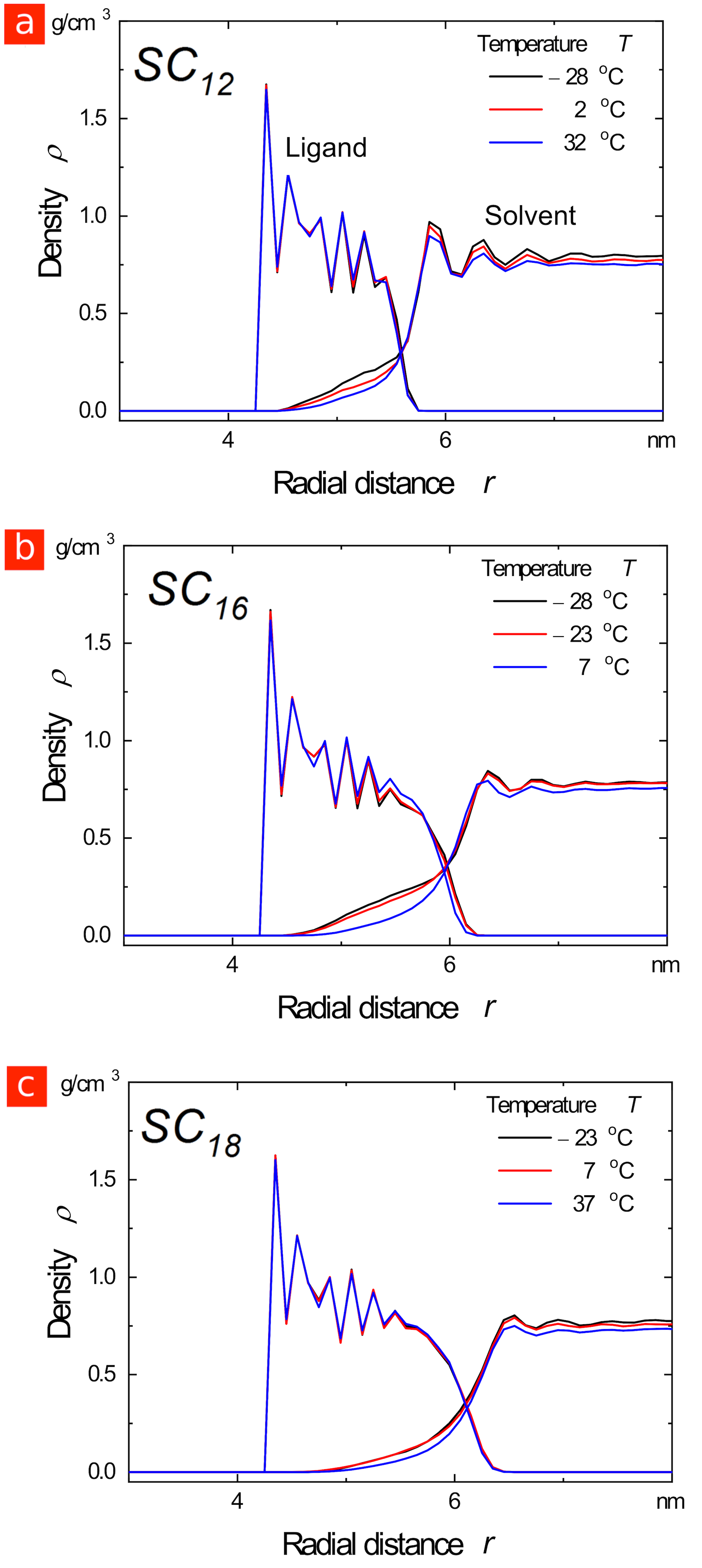}
    \caption{Radial density distributions for the decane solvent and for alkanethiol ligands of different lengths on \SI{8.3}{\nano\meter} gold cores of AuNP, plotted as a function of the distance \(r\) from the center of the nanoparticle core. The red lines indicate the density profile at the ligand ordering temperature \(T_\mathrm{order}\) for each particle.}
    \label{SI_DensityProfile_8nmAu}
    \vspace{0.5em}
\end{figure}

Figure \ref{SI_DensityProfile_8nmAu} shows how the density of the solvent and different length ligands changes as we move away from the \SI{8.3}{\nano\meter} gold core. The ligand-solvent interface is more sharply defined for shorter ligands, and the ordering of the ligands that occurs at low temperature allows more solvent to enter the ligand shell.

\begin{figure}[H]
    \centering
    \includegraphics[width=0.9\linewidth]{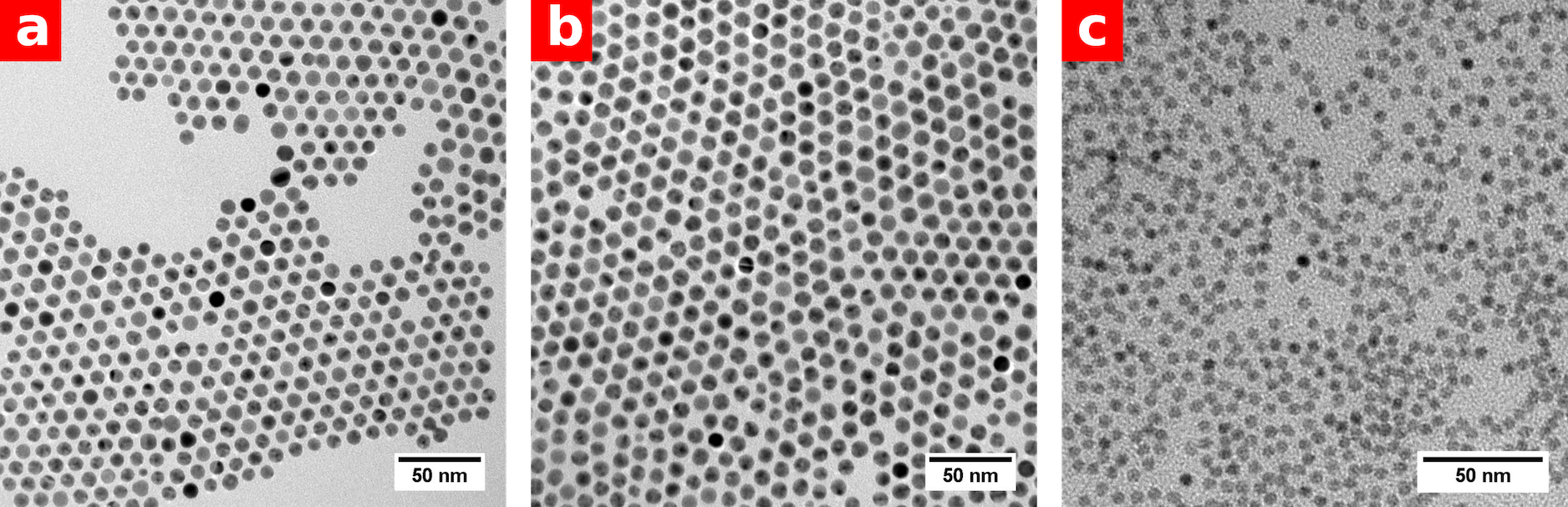}
    \caption{Transmission electron microscopy micrographs of AuNP with a diameter of (a) \SI{8.3}{\nano\meter}, (b) \SI{8.9}{\nano\meter} and (c) CdSeNP with a diameter of \SI{6.0}{\nano\meter}.}
    \label{SI_TEM}
    \vspace{0.5em}
\end{figure}

Figure \ref{SI_TEM} shows TEM data of the as-synthesized, oleylamine-capped Au and CdSe nanoparticles. The particles are roughly spherical in shape, but show faceting, even at low resolutions, especially for AuNP.

\begin{figure}[H]
    \centering
    \includegraphics[width=0.8\linewidth]{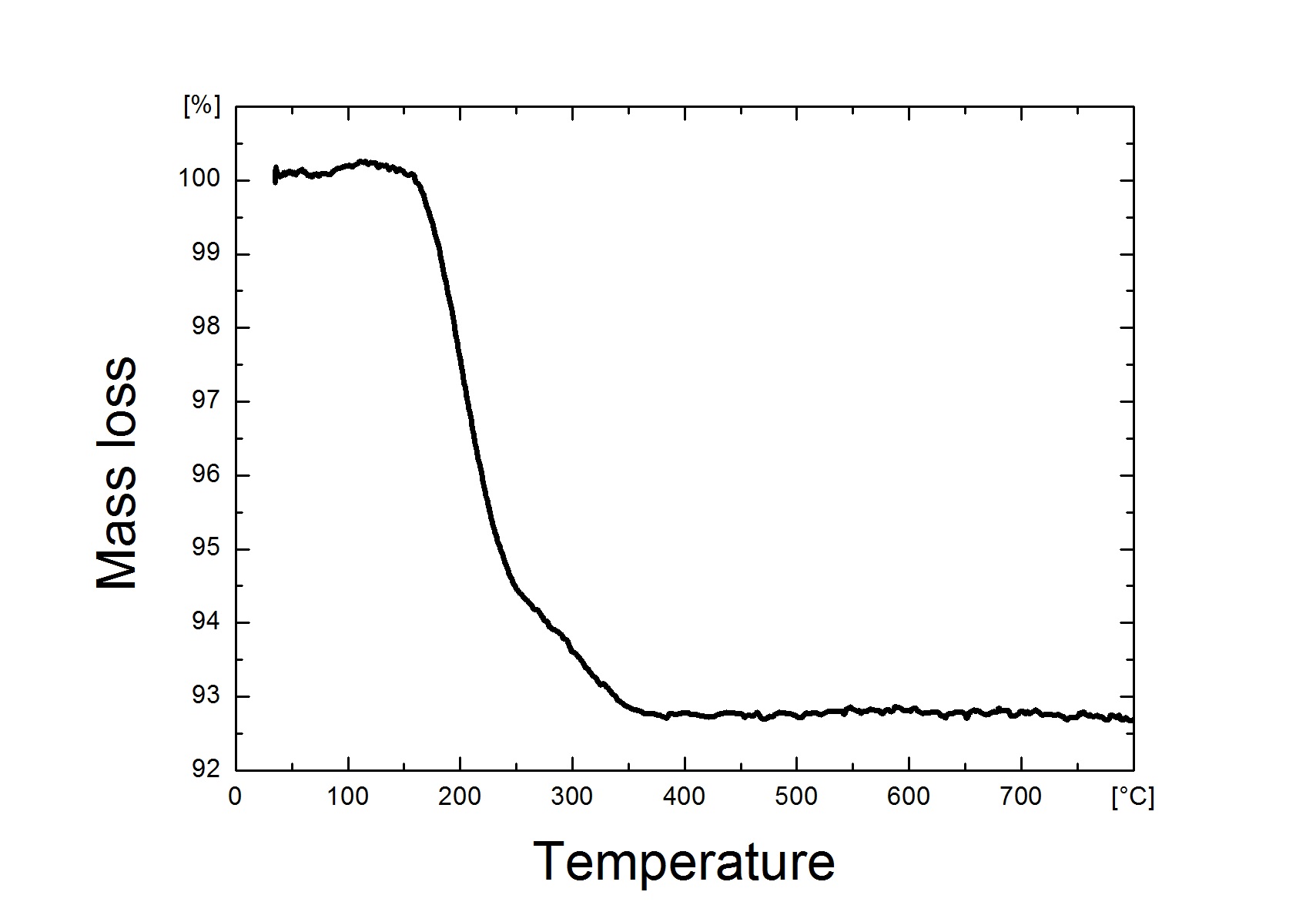}
    \caption{\Revised{Thermogravimetric analysis of 1-hexandecanethiol coated AuNP with a diameter of \SI{8.9}{\nano\meter}. Ligand degradation started a around \SI{150}{\celsius} and finished at \SI{400}{\celsius}.}}
    \label{SI_TGA_Au}
\end{figure}

\Revised{Figure \ref{SI_TGA_Au} shows a thermogravimetric analysis of 1-hexandecanethiol coated AuNP with a core diameter of \SI{8.9}{\nano\meter}. A mass loss of around \SI{7.3}{\percent} is due to the ligand that is fully removed at around \SI{400}{\celsius}.}

\begin{figure}[H]
    \centering
    \includegraphics[width=0.8\linewidth]{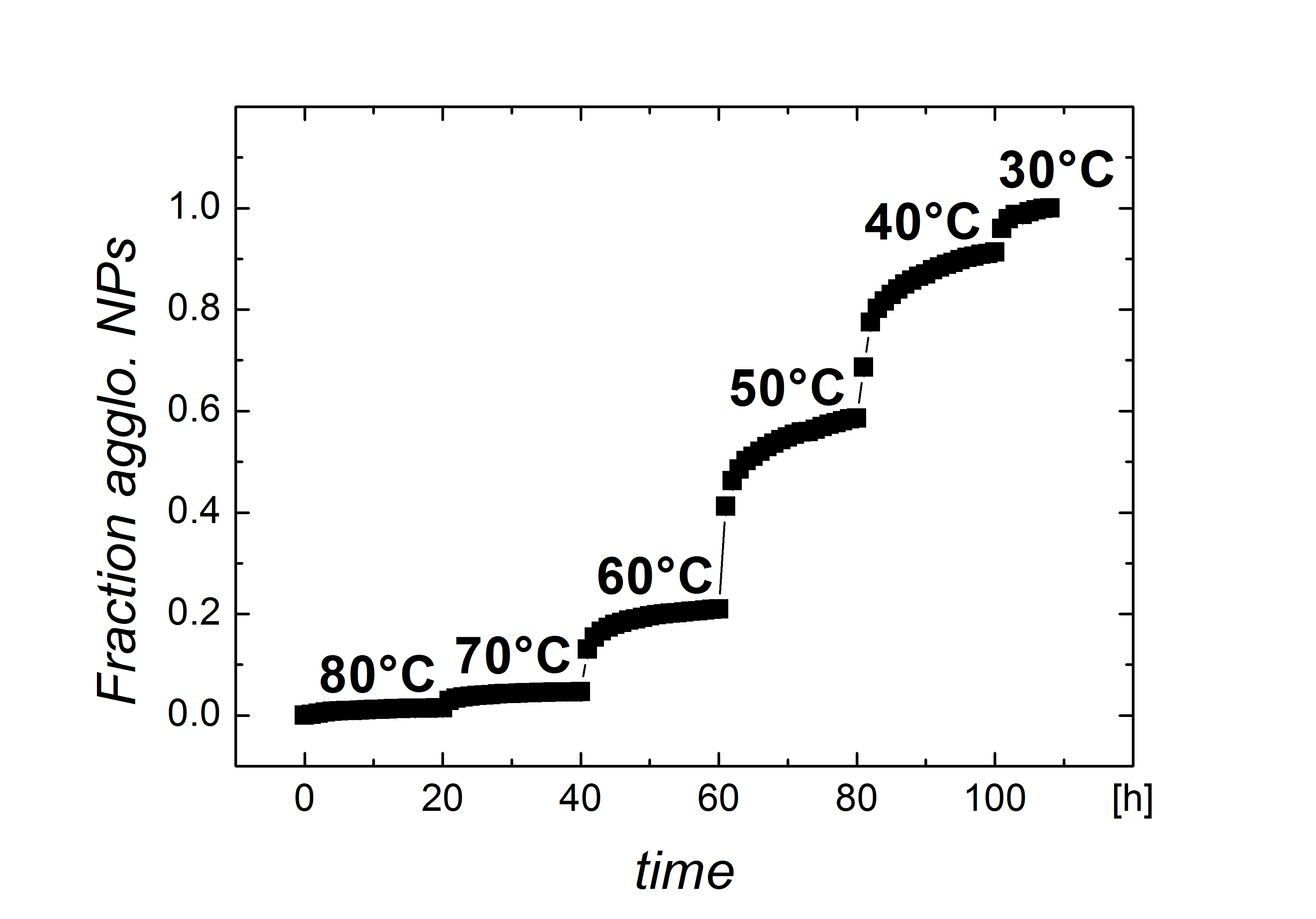}
    \caption\Revised{{Transient agglomeration data on 1-hexadecanethiol coated AuNP with a core diameter of \SI{8.3}{\nano\meter}. The fraction of agglomerated NPs was measured every \SI{30}{\minute}. The sample was kept for \SI{20}{\hour} at each temperature.}}
    \label{SI_SAXS_time}
\end{figure}

\Revised{Figure \ref{SI_SAXS_time} shows the time-dependent agglomeration of 1-hexadecanethiol coated AuNP with a core diameter of \SI{8.3}{\nano\meter}. Such experiments were performed to ensure that the largest fraction of particles that lost colloidal stability at a given temperature had agglomerated. To this end, the fraction of agglomerated NPs was measured every \SI{30}{\minute}, beginning at \SI{80}{\celsius} with a total waiting time of \SI{20}{\hour} for each temperature.}

\begin{figure}[H]
    \centering
    \includegraphics[width=0.8\linewidth]{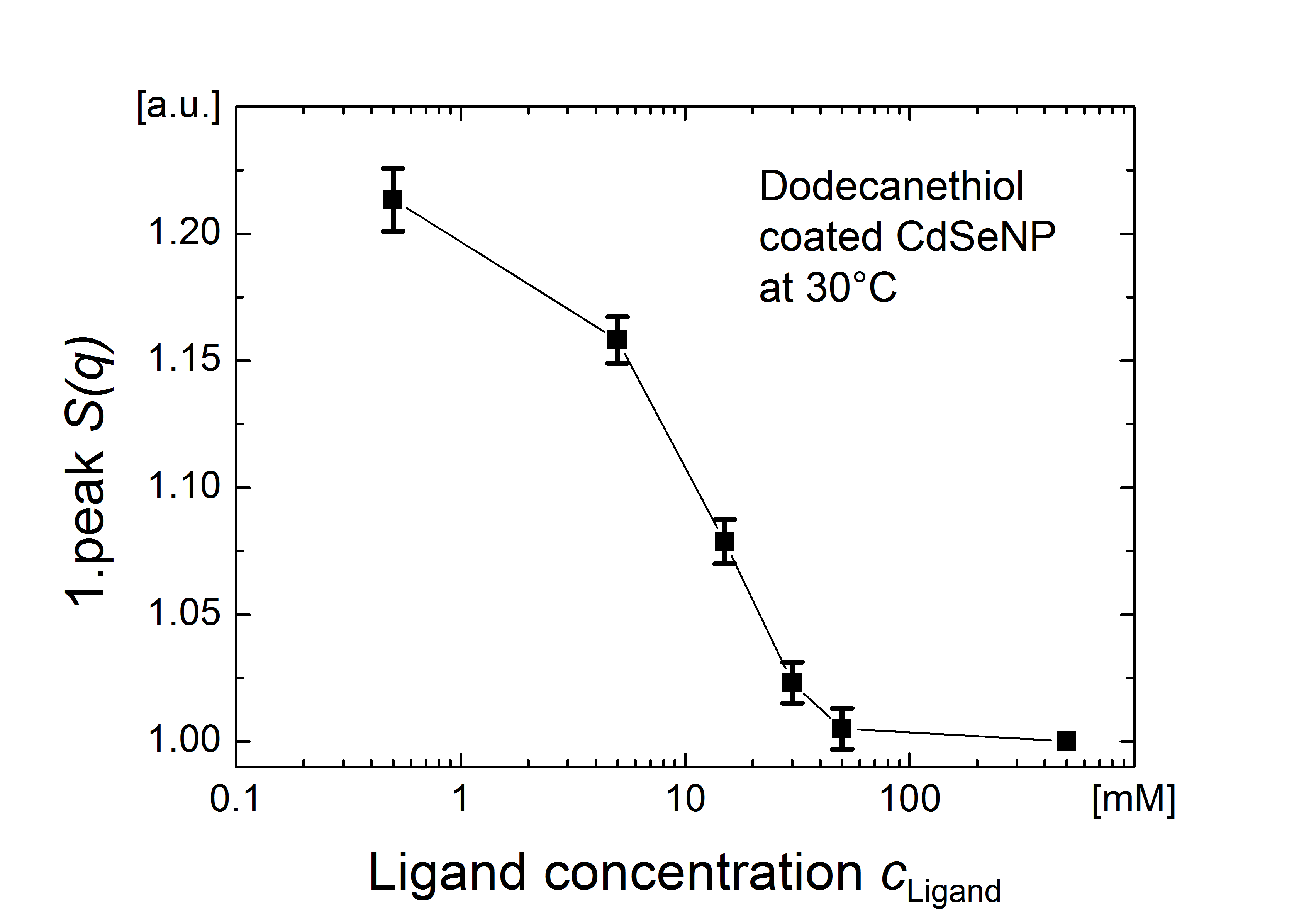}
    \caption{Change in the position of the first peak of the structure factor S(q) with dodecanethiol ligand concentration \(c_{Ligand}\) for CdSeNP suspensions in decane at \SI{30}{\celsius}.}
    \label{SI_CdSeLigConc}
\end{figure}

\begin{figure}[H]
    \centering
    \includegraphics[width=0.8\linewidth]{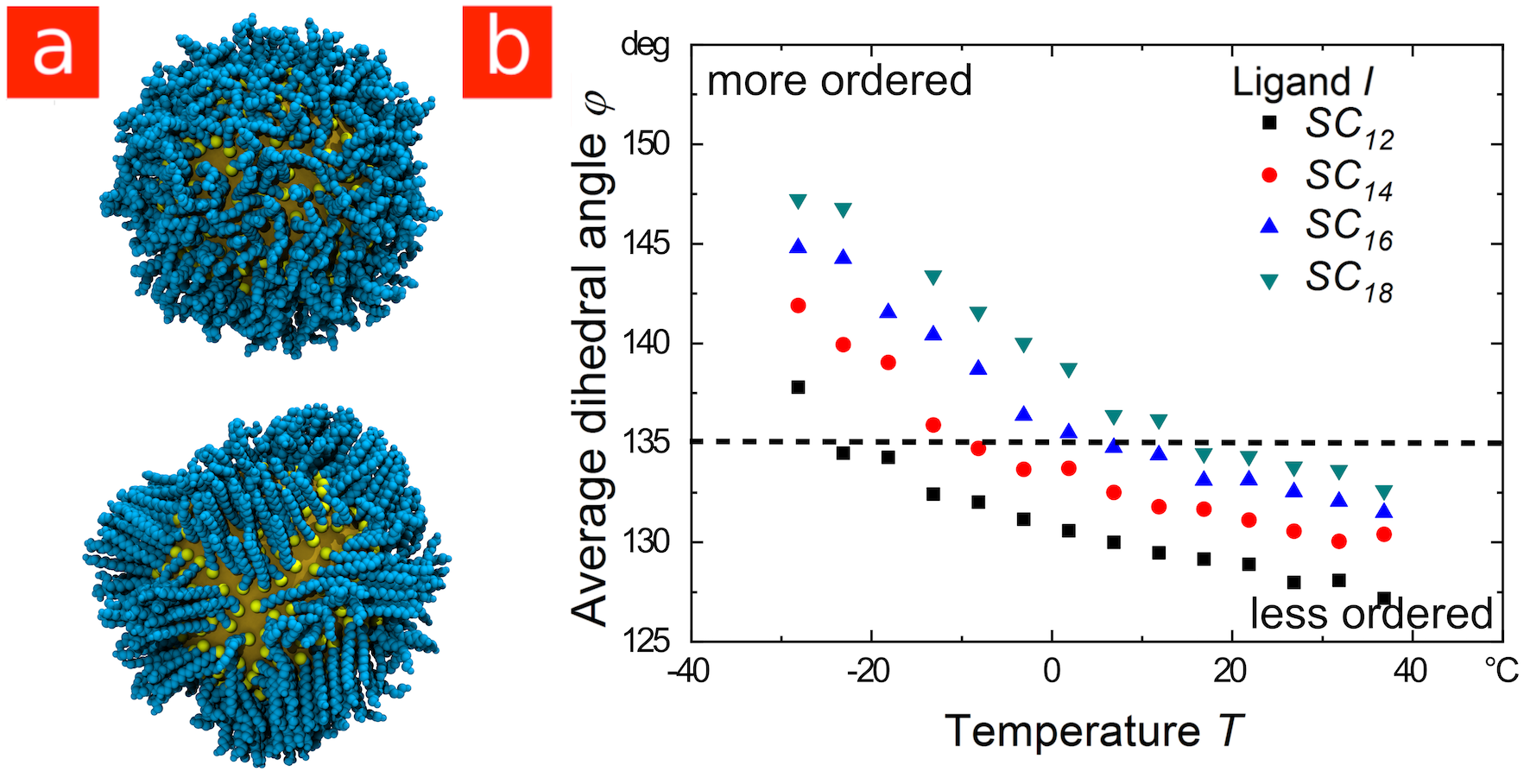}
    \caption{(a) Snapshots of the simulations of \SI{5.8}{\nano\meter} CdSe particles coated with \(SC_{16}\) ligands at a surface coverage of \SI{3.6}{\nano\meter^{-2}} at \SI{57}{\celsius} (top) and \SI{-28}{\celsius} (bottom). Even at low coverage, the ligands extend in bundles on the surface of the particle towards the decane solvent. (b) Average dihedral angle for the ligands. The transition from less to more ordered is broader and shifted down by approximately \SI{40}{\celsius} compared to when the surface coverage is \SI{5.5}{\nano\meter^{-2}}.}
    \label{SI_CdSeLowCoverage}
    \vspace{0.5em}
\end{figure}

\begin{figure}[H]
    \centering
    \includegraphics[width=0.9\linewidth]{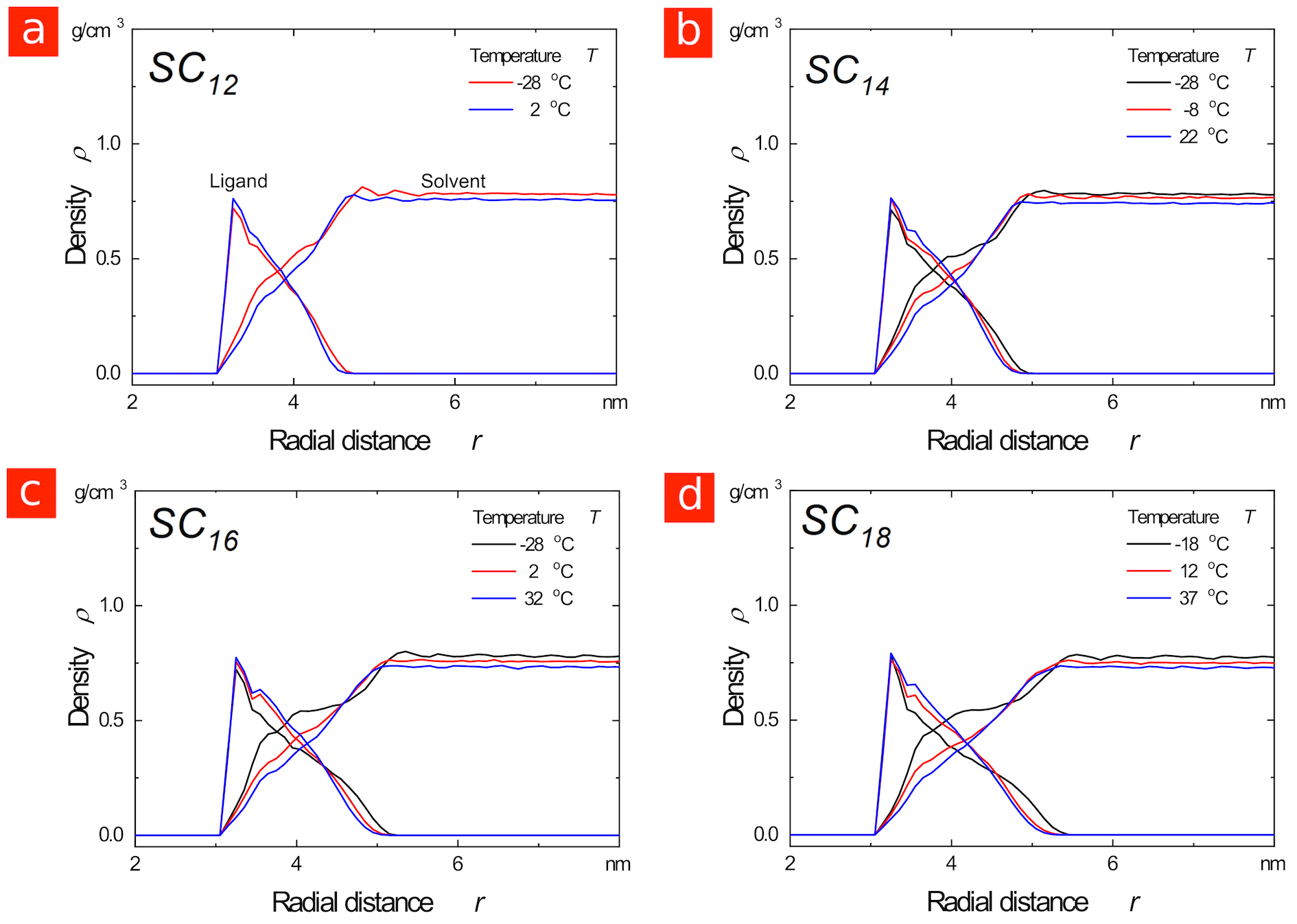}
    \caption{Radial density distributions for the decane solvent and for ligands of different lengths coating \SI{5.8}{\nano\meter} CdSeNP, at a surface coverage of \SI{3.6}{\nano\meter^{-2}}, plotted as a function of the distance \(r\) from the center of the nanoparticle core. The red lines indicate the density profile at the ligand ordering temperature \(T_\mathrm{order}\) for each particle.}
    \label{SI_density}
    \vspace{0.5em}
\end{figure}

Figure \ref{SI_CdSeLigConc} shows how CdSeNP suspensions are affected by the addition of excess ligand. In decane, a fully dispersed state was only achieved at a ligand concentration of \SI{500}{\milli\Molar}, indicating the adsorption/desorption equilibrium of alkanethiol ligands on these particles. This suggests that the surface coverage might vary noticeably for CdSeNP, as opposed to what was observed for AuNP\cite{Kister2018}. As discussed in the main manuscript, we therefore performed additional simulations at a lower coverage of \SI{3.6}{\nano\meter^{-2}} to obtain results reported in Figures \ref{SI_CdSeLowCoverage} and \ref{SI_density}.

\bibliography{bib}